\documentclass[twoside]{article}
\usepackage{PRIMEarxiv}
\usepackage[utf8]{inputenc} 
\usepackage[T1]{fontenc}    
\usepackage{hyperref}       
\usepackage{url}            
\usepackage{booktabs}       
\usepackage{amsfonts}       
\usepackage{nicefrac}       
\usepackage{microtype}      
\usepackage{lipsum}
\usepackage{fancyhdr}       
\usepackage{graphicx}       
\usepackage{amssymb}

\title{A Novel Cloud-Based Framework for Standardised Simulations in the Latin American Giant Observatory (LAGO)
\thanks{\textit{\underline{Citation}}: 
\textit{A. J. Rubio-Montero, R. Pagán-Muñoz, R. Mayo-García, A. Pardo-Diaz, I. Sidelnik and H. Asorey, A Novel Cloud-Based Framework For standardised Simulations In The Latin American Giant Observatory (LAGO), 2021 Winter Simulation Conference (WSC), 2021, pp. 1-12, doi: 10.1109/WSC52266.2021.9715360.}\\
{The LAGO Collaboration. See the complete lists of authors and institutions at \href{http://lagoproject.net/collab.html}{http://lagoproject.net/collab.html}}}
}
\author{Antonio Juan Rubio-Montero, Raúl Pagán-Muñoz, Rafael Mayo-García\\
Centro de Investigaciones Energéticas,\\
Medioambientales y Tecnológicas (CIEMAT)\\
Av. Complutense 40,
Madrid, 28040, SPAIN\\
\And
Alfonso Pardo-Diaz\\
Centro Extremeño de Tecnologías\\
Avanzadas (CETA-CIEMAT) \\
Calle Sola 1,
Trujillo, 10200, SPAIN
\And
Iván Sidelnik\\
Departamento de Física de Neutrones\\
Centro Atómico Bariloche (CNEA/CONICET),\\
San Carlos de Bariloche, 8400, ARGENTINA\\
\And
Hernán Asorey\\
Instituto de Tecnologías en Detección y\\
Astropartículas (ITeDA, CNEA/CONICET/UNSAM)\\
Centro Atómico Constituyentes,\\ 
Villa Maipú, 1450, ARGENTINA
}

\begin{document}
\maketitle

\begin{abstract}
LAGO, the Latin American Giant Observatory, is an extended cosmic ray observatory, consisting of a wide network of water Cherenkov detectors located in 10 countries. With different altitudes and geomagnetic rigidity cutoffs, their geographic distribution, combined with the new electronics for control, atmospheric sensing and data acquisition, allows the realisation of diverse astrophysics studies at a regional scale. It is an observatory designed, built and operated by the LAGO Collaboration, a non-centralised alliance of 30 institutions from 11 countries.

While LAGO has access to different computational frameworks, it lacks standardised computational mechanisms to fully grasp its cooperative approach. The European Commission is fostering initiatives aligned to LAGO objectives, especially to enable Open Science and its long-term sustainability. This work introduces the adaptation of LAGO to this paradigm within the EOSC-Synergy project, focusing on the simulations of the expected astrophysical signatures at detectors deployed at the LAGO sites around the World.
\end{abstract}

\section{Introduction}\label{sec:intro}

Astroparticle physics is nowadays one of the scientific fields that evidences large interdisciplinary contributions, given the variety of topics that this discipline covers: from high energy astrophysics (e.g., acceleration mechanisms at astrophysical objects such as supernovae remnants) to computational sciences (as this work shows).

\begin{figure}[!ht]
  \centering
  \includegraphics[scale=0.65]{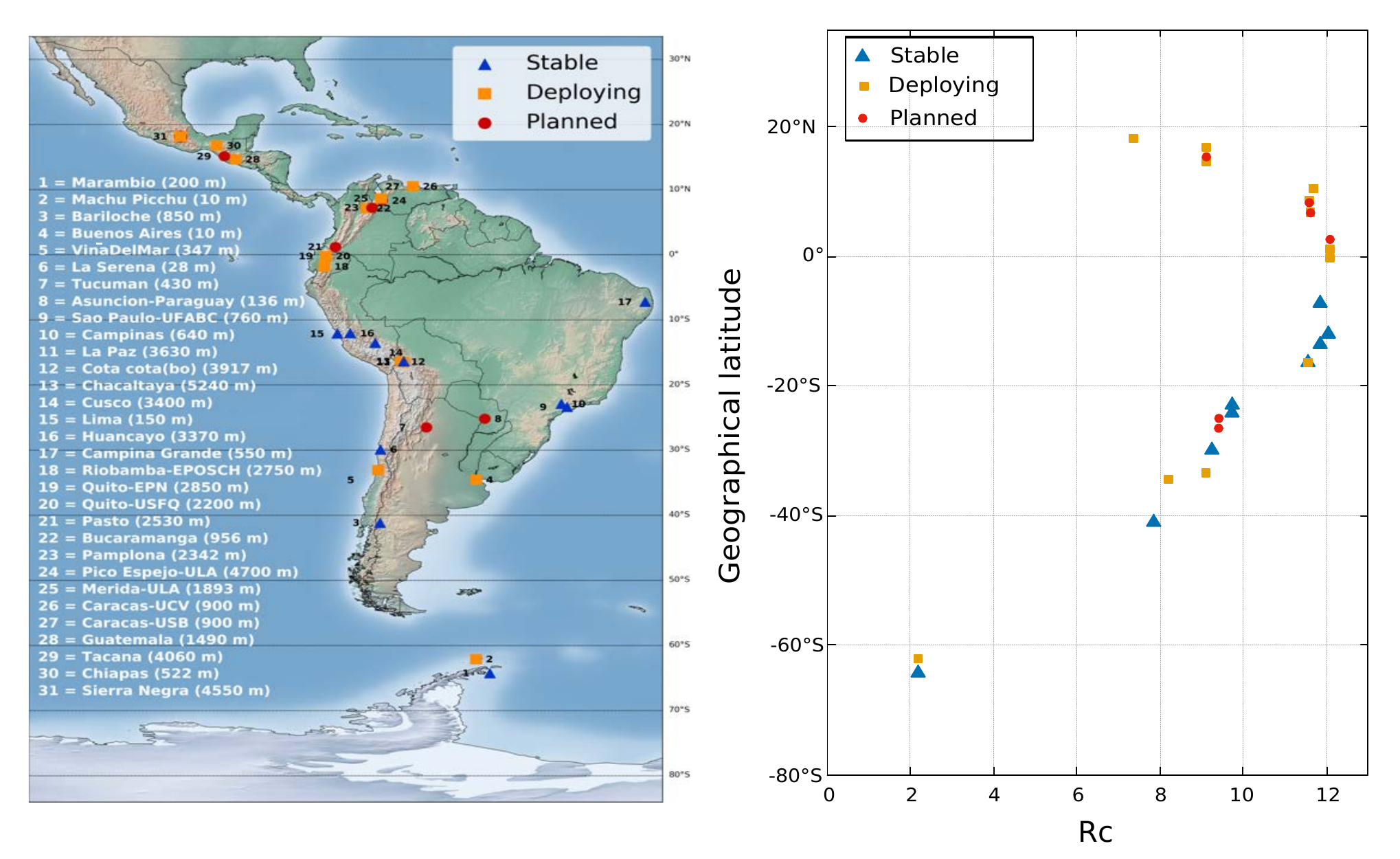}
  \caption{(left) Geographical distribution and altitudes of the LAGO water Cherenkov detectors: the ones in operation are represented with blue triangles, orange squares are used for those detectors that are being deployed, and the planned sites are indicated with red circles.
    (right) Vertical geomagnetic rigidity cutoff at each LAGO site.}
  \label{fig:figlagomap}
\end{figure}

The Latin American Giant Observatory (LAGO) is a project originally conceived in 2006\,\cite{Allard2008} to detect the high energy component of Gamma Ray Bursts (GRBs), with typical energy of primaries $E_p \gtrsim 20$\,GeV. It consisted of installing water Cherenkov detectors (WCDs) at high altitude mountains across the Andean ranges.
From this initial aim, LAGO has evolved towards an extended astroparticle observatory at a Continental scale by operating WCDs and other particle detectors in ten countries in Latin America (LA).
The current network spans a region from the south of Mexico, with a small array installed at $4,550$\,m a.s.l.\ in Sierra Negra\,\cite{Galindo2017}, to the Antarctica Peninsula\,\cite{Dasso2019}.
As can be seen in Figure~\ref{fig:figlagomap}, the LAGO detection network is distributed over similar geographical longitudes but a wide range of latitudes and altitudes are covered, being able to compile data for a large range of geomagnetic rigidity cutoffs and atmospheric absorption depths\,\cite{Sidelnik2015}.

The WCDs of LAGO are composed of a typically cylindrical $1-4$\,m$^3$ water container with a highly reflective and diffusive coating inside.
In the centre of the container top, a photomultiplier tube (PMT) is placed pointing downward to convert the Cherenkov radiation produced by the interaction of high energy particles with the water volume into an electrical signal.

The LAGO detection method is an implementation of the single-particle technique (SPT)\,\cite{Vernetto2000} that perform its studies by looking for significant deviations in the counting of background signals at different sites.
While this technique lacks of directional reconstruction, it is however possible to use atmospheric absorption as a selection tool for periodic signals including gamma point sources.
So, by performing simultaneous measurements of the modulated flux of the cosmic rays at different locations covering a wide range of geomagnetic rigidities and atmospheric reactions, LAGO can provide important information about high-energy astrophysical transients and space weather phenomena, even in an almost real-time basis.

If enough computational power is available, it is possible to simulate the complete response of the system, including the geomagnetic, atmospheric and detector conditions, to gain a proper understanding of the signals flux registered at any detector.
These simulations also enable the estimation of the measurements of a wider WCD network, exploring new locations for WCDs or developing new calibration methods.

For doing this, the Collaboration needs to count on standardised computational mechanisms for profiting the simulation framework that has been developed\,\cite{Asorey2018,Sarmiento2019}, with special emphasis on the coordination, share, and curation of the results.
In this sense, the European Commission is fostering multiple initiatives aligned to LAGO objectives, specially to adopting the Open Science paradigm as a way to guarantee the long-term sustainability of the research.
These initiatives are based on observing FAIR (Findable, Accessible, Inter-operable and Re-usable data) principles\,\cite{Wilkinson2016}, the federation of virtual resources (cloud services and virtual organisations) and virtual observatories.

In this work, we introduce the framework developed within the EOSC-Synergy\,\cite{EOSC} project, which will be considered as the new standard for simulations in the LAGO Collaboration.

\section{Simulations in LAGO}\label{sec:sc_lago}

LAGO is mainly oriented to perform basic research focusing on three main areas: high energy phenomena\,\cite{Asorey2015b}, the measurement of atmospheric radiation at ground level\,\cite{Galindo2017} and space weather and climate monitoring\,\cite{Asorey2015a,Asorey2018}.

Due to its social and economic impact, space weather is now one of the main areas of space research.
As it has been well established, transient solar ejecta and changes in the global structure of the heliospheric and terrestrial magnetic fields are closely related, and they can inter-modulate the flux of cosmic rays of galactic origin (GCRs).
So, the measurement of the temporal variations of this flux provides information that can be related to the transport of particles in the inner and outer heliosphere.
While long-term modulations (months to years) in the flux of GCR are associated with the solar cycle, the short-term ones in periods from hours to weeks are produced by interplanetary transient perturbations in the calm solar wind causing phenomena like the Forbush Decreases (FD)\,\cite{Cane2000}.
Observed from ground level, FDs exhibit asymmetrical structures: fast decreases (several hours) of the cosmic ray (CR) flux are followed by smooth recoveries (several days).

Our observations have demonstrated that, by using adapted analysis techniques to the characteristics of our small detectors, it is possible to observe space weather phenomena at different time scales from ground level in the LAGO network of WCD\,\cite{Asorey2015a}.
As mentioned in the section\,\ref{sec:intro}, by harnessing the advantages of the WCDs,
it is possible to determine the flux of secondary particles at different bands of deposited energy within the detector volume.
These bands are dominated by different components of the secondary particles produced during the interaction of GCR with the Earth's atmosphere.
Those are the electromagnetic (photons and electrons), muonic and hadronic components.
This multi-spectral analysis technique (MSAT) constitutes the basis of our Space Weather oriented data analysis\,\cite{Asorey2015a}.
By combining all the data measured at different locations of the detection network, the LAGO project will provide detailed and simultaneous information on the temporal evolution and small- and large-scale characteristics of the disturbances produced by different transient and long-term space weather phenomena, some of them even in an almost real-time basis.

A complex simulation sequence involves all the related aspects of the physics of LAGO: the calculation of the expected flux of signals at any site in the World;
the design and development of new detectors and data analysis techniques;
and even the development of new experiments to be conducted in our lectures at the participating Universities\,\cite{Asorey2018a}.

\subsection{ARTI: the LAGO Simulation Framework}\label{subsec:compu}

Each WCD of LAGO is designed and modelled to estimate its responses to cosmic background radiation according to its geographical position.
To this end, the LAGO Collaboration has developed ARTI\,\cite{Sarmiento2019}, a framework of codes focused on estimating the cosmic ground radiation flux and the detectors' response to it, by joining:
the Global Data Assimilation System (GDAS) to get the physical atmospheric profiles\,\cite{gdas2021}; MAGCOS, application to simulate the evolving conditions of the geomagnetic field\,\cite{Desorgher2003};
CORSIKA,
a complete Monte Carlo code to simulate the interaction of astroparticles with the atmosphere\,\cite{Corsika}; and, finally, the Geant4
toolkit\,\cite{Agostinelli2003} to get the expected response of the LAGO detectors.

To do this, the first step is to calculate the expected number of primaries by integrating the measured flux ($\Phi$) of each GCR species ($1<Z<26$) within the $1<E_p/\mathrm{GeV}<10^6$ energy range in $1$\,m$^2$ of atmosphere at an altitude of $112$\,km a.s.l.\
Here, $\Phi$ is considered as:
\begin{equation}
 \Phi(E_p, Z, A, \Omega) \simeq j_0(Z,A)
 \left(\frac{E_p}{E_0}\right)^{\alpha(E_p,Z,A)}\, ,\label{eq:equation}
\end{equation}
where $E_p$ is the energy of the particle, $\alpha (E_p, Z,A ) $ is considered constant with energy, i.e., $\alpha(E_p,Z,A)\simeq\alpha(Z,A)$, from $10^{11}$\,eV to $10^{15}$\,eV and $E_0$ has a value of $10^{12}$\,eV\@.

To account for the geomagnetic effects, at every LAGO site a tensor containing the directional rigidity cutoff $R_C$ of the site is calculated including the local conditions of the geomagnetic field at this moment, by considering secular and disturbed geomagnetic field conditions, such as those produced during intense geomagnetic storms.
Geomagnetic field corrections are calculated by using the 13$^\mathrm{th}$ generation of the International Geomagnetic Reference Field (IGRF13-2019) and Tsynganekov (TSY2001) models\,as described in~\cite{Asorey2018}.

As cosmic rays will produce a cascade of secondary particles during their interaction with the atmosphere, a proper knowledge of local atmospheric conditions is needed, and so, the local atmospheric profiles are extracted from the Global Data Assimilation System (GDAS)\,\cite{gdas2021}.

Then, the full-sky expected number of primaries $\Phi$ at each site is simulated by using CORSIKA to determine the expected number of secondary particles at ground level.
For a $1$\,m$^2$ detector and an integration time of $1$\,hour, a typical run per site may require the simulation of $\sim (2-5) \times 10^7$ primaries, depending on the minimum rigidity cutoff of the site ($\lesssim 2$\,GV for the LAGO antarctic sites) and produce $\sim 25$\,GB of compressed data.
The final set of secondary particles at the ground is obtained by comparing the primary rigidity $r(p,Z)=pc/Z$, where $p$ is the primary momentum and $Z$ its charge, with the tensor containing the local geomagnetic cutoffs.
By doing this, only those secondaries that hit the ground and were produced only by allowed primaries are selected.
This method has the advantage of allowing the usage of the same site simulation set in the evolving conditions of the geomagnetic field, see~\cite{Asorey2018} for more details.

Finally, as described in~\cite{Sarmiento2019} the selected secondary particles are injected in the Geant4-based simulated detector.
The model considers the detector materials and geometry, its inner coating, the PMT model (characterised by its geometry, quantum efficiency to different photon wavelengths, electronics, and time response) and even the water quality and the presence of different solutes.
The Cherenkov photons produced by the secondary particles in the water volume are propagated through the detector and finally, the expected pulses are obtained and stored.
The final results are the time series of the expected flux of secondaries and signals at the detector, and the simulated charge histograms that contains information about the deposited energy in the detector.
These figures are important for the detector calibration, as the deposited charge in the PMT for a central and vertical muon (VEM) going through the detector is the main calibration parameter and depends on many factors, such as the water and coating quality, the detector geometry, the detector trigger or the PMT polarisation voltage.

\subsection{Hierarchical Model for the Simulated Data} \label{subsec:hieralchical-model-for-the-simulated-data}

With the ARTI framework it is possible to obtain the expected flux of signals at a specific detector in a particular site, and under realistic atmospheric and geomagnetic time-evolving conditions.
Thus, researchers are able to generate different types of hierarchical data-sets in a pipelined fashion:
\begin{itemize}
  \item S0 Plain simulations: CORSIKA raw simulated data;
  \item S1 Analysed simulations: ARTI analysis of the S0 data-set, corresponding to the expected flux of secondary particles at ground level; and
  \item S2 Detector simulations: Geant4 simulation of the S1 data-set with the time-series of the expected detector response signals, and the calibration charge histograms.
\end{itemize}

\section{Standardising LAGO Computations}\label{sec:standardising-lago-computations}

\subsection{Precedents and Open Science}\label{subsec:precedents-and-open-science}

From the beginning of the LAGO Collaboration, several threats were identified related to the management of the generated data and shared resources:
i) the lack of computational and storage resources having a common accessibility method and without compatibility issues among platforms;
ii) the absence of effective mechanisms for the curation and sharing of the generated data;
and iii) the difficulty of coordinating the development, tracking and deploying the official releases of codes.

Several initiatives were proposed to face these problems, achieving a different degree of success.
The lack of resources was mitigated with grid computing techniques, not solving the compatibility issues, and the ease of use.
Within LAGO, storage and cluster facilities were offered by some institutions\,\cite{rodriguez2015res},
but an unified access was not yet established.
To overcome these issues, an IaaS cloud federation was initially tested\,\cite{Mocskos2018}, but there was a dependency on proprietary orchestrators.
Additionally, the middleware running on the WCDs\,\cite{Asorey2015f} was using plain metadata, not following any known schema, which becomes unfeasible to be referenced at a later stage, even by other LAGO applications, such as ARTI. A service for catalogues was developed\,\cite{Torres2011} to expose consolidated results, however, it was unsuitable for making deep mining into the metadata or data.
To preserve the version control of the developed tools and promote collaboration among developers, the GitHub hosting platform was employed, but the potential of its advanced features was not fully used on the coordinating activities, to assure the quality of codes and documentation.

So, individual initiatives were not enough to overcome the these limitations.
The complexity and size of LAGO made necessary a joint approach, which can be framed within the Open Science paradigm.
Yet, Latin America is a preferred scientific partner for the European Commission (EC).
To assure the continuity of research programs, the EC promotes Open Science and always includes a policy priority to standardise the scientific methods and processes of its funded research and innovation programs.
The main idea behind Open Science is to make the investigations more transparent and the results more accessible, helping to disseminate the latest developed knowledge.
Open Science is the practice of science in such a way that others can collaborate and contribute, where research data, lab notes and other research processes are freely available, under terms that enable reuse, redistribution, and reproduction of the research and its underlying data and methods\,\cite{Bezjak2018}.

Regarding Open Science, two main ambitions of the EC are associated to the LAGO objectives: Open Data and European Open Science Cloud (EOSC).
Open Data relates to the application of what is known as FAIR principles for all results of EC-funded scientific research: Findable, Accessible, Inter-operable and Re-usable data.
While EOSC is a trusted virtual environment for hosting and processing research data to support EC science, that cuts across borders and scientific disciplines to store, share, process, and reuse research digital objects complying the FAIR principles\,\cite{EOSC}.
Therefore, the objectives of EC and LAGO are aligned, since the final purpose of both is to enable the global profit and contribution of this research, within and outside the LAGO Collaboration, through a sustainable virtual observatory and a standardised computational model.

\subsection{Requisites and Constraints}\label{subsec:requisites-features-and-constraints}

The adoption of the Open Science paradigm is a laborious design process that includes: the capture of requirements;
the definition of data and metadata, and the mechanisms of data generation, preservation, and publication;
an open discussion on the adoption of suitable technologies and future trends;
and finally, putting all together in a common document called the Data Management Plan (DMP), which becomes the standardisation guide for the project development.
Concerning the requirements, some threats were already described above.
Besides, the idiosyncrasy of LAGO should also be considered.
The following issues about the collaborators are of importance:
\begin{itemize}
  \item They are grouped into autonomous research units within different work packages, with every unit having specific responsibilities.
  For example, every detector has an operator unit, every software piece has a manager, etc.
  The external staff is allowed in or removed from the collaboration by each research group for specific contributions.
  \item Most of the involved people are students or researchers in astrophysics and HEP with a solid background in computing skills, i.e., they are accustomed to profiting from HPC facilities and/or using control version systems such as Git.
  \item Results of every contributor (simulations, processing data, curating measurements, coding, etc.) are of interest of other members or external actors.
  The generated data or codes should be protected and registered, but also shared and published after an embargo period.
  \item Some institutions support the project by sharing their computational resources, such as clusters, and their associated storage, but they are generally non-exclusive, providing only a limited environment.
\end{itemize}

In addition, the LAGO community also imposes additional constraints to the adoption of Open Science within EOSC: the independence of LAGO must be preserved, it cannot be completely tied to EOSC services;
the self-sustainability of the new computing model, deploying services in or out EOSC;
and the selected standards, protocols, and technologies should foresee the continuity of the project.

As previously indicated, the\,\cite{LAGO-DMP} is the main resource, supporting and integrating all the requirements for implementing the Open Science paradigm in LAGO\@.
This tool is the essential reference to assure the FAIR principles of the results and, among other things, establishes the format of the data and the metadata and the protocol for generating, storing, and accessing the results.
DMPs should be living documents that will be amended, improved and detailed along the project timeline.
For this purpose, the\,\cite{LAGO-DMP} for LAGO is managed at GitHub to maintain clear versioning.

\subsection{Schemes for Data and Metadata}\label{subsec:schemas-for-data-and-metadata}

The consolidated trend in Open Data is making use of standardised schemes based on linked data.
LAGO has followed this direction when selecting the definitions for the metadata of the project, as well as because of its simplicity and compatibility:

\begin{itemize}
  \item \textbf{Language syntax} JSON-LD 1.1 (W3C).
  Being one of the simplest standard for linking metadata.
  Promoted by Google, it is now ousting the heaviest syntaxes such as RDF/XML or Turtle.
  \item \textbf{Main vocabulary} DCAT-AP2 (European Commission), a specific profile of DCAT2 (W3C), recommended for repositories, content aggregators or data consumers related to the public sector (government, research centres, funded projects).
  \item \textbf{The LAGO vocabulary} described in this document~\cite{lago-metadata}, is a re-profile of DCAT-AP2, extending the existing classes and adding properties needed for the LAGO computation, such as describing geomagnetic locations, expected atmospheres or software used.
\end{itemize}

Concerning the data organisation itself, they are structured in catalogues and datasets, but with certain considerations.
Each generated file is considered the minimum dataset to be data-linked and processed, while a collection of related files is grouped in a catalogue, which should be referenced with a persistent and unique identifier (PiD).
In this sense, Handle.net PIDs will never point directly to datasets (i.e., files), but to catalogues or repository providers.
The former are used as shortcuts for publication (as cheaper DOIs), but the latter are part of the IRIs. As resolvable IRIs (IETF RFC 3987) univocally identify every JSON-LD object (datasets, catalogues, and others) over the Internet, any dataset (or even a catalogue) could be referenced by its absolute IRI, composed by its provider PID, its internal path and its name.
Therefore, a dataset does not need an exclusive PiD, since the different LAGO activities generate only one data sub-type.
catalogues will only contain files belonging to the sub-type activity, and catalogues are adequate for general reference.
Regarding the naming conventions, the catalogue name will be based on the data type (S0-S2) by adding meaningful parameters used by the software, separated by an underscore, ``\_'', e.g., $S0\_<site>\_<flux time>\_<altitude>\_<other\ params>$.
By combining this naming and the IRI mechanisms, the accessibility to the data through provider migrations (changes of name or domain of storage) can be assured, as well as referencing concrete files, if needed.
The preferred way to state this feature in JSON-LD is by using the @base property in the @context, becoming whole IRIs relative to the @base PiD\@.

\subsection{Selected EOSC Services}\label{subsec:selected-eosc-services}

LAGO has already integrated the following services from the EOSC marketplace through its standardisation:

\begin{itemize}
  \item \textbf{EduTeams} is an identity provider (IdP) that is associated with a Perun service, both supported by GEANT. This enables the federation of identities specifically for research communities, i.e., the creation of virtual organisations (VOs).
  Managing the LAGO VO with Perun at GEANT was considered because of its better flexibility over EGI Check-in IdP (see below), and because it has a certain independence from EC-related projects and longer-terms support to Latin American users.
  Perun provides the needed flexibility allowing several roles and permissions over the data, such as conventional users (allowing them to see a whole data-set but with writing restrictions), research group chiefs (allowing them to enrol new researchers), robots, main administrators, etc.
  Yet, the sustainability of the VO in Latin  America is guaranteed by the support of RedClara (associated with GEANT), allowing to extend users and resources beyond EOSC\@.
  \item \textbf{EGI Check-in} is the prevalent IdP in EOSC, thus it is needed for accessing most of its services, such as described below.
  EGI Check-in also supports external IdPs, acting as a router of EduTeams identities and being transparent to LAGO users.
  \item \textbf{EGI DataHub} is one of the cloud storage services available in EOSC. It is based on OneData, a globally distributed cloud file-system that allows researchers several ways to access the data and metadata of their interest.
  Collaboration members can directly explore the directory tree at https://datahub.egi.eu, mount it on their PC's, or make direct use of REST APIs. Meanwhile, the public will search the published results on external repositories, which are harvesters of the metadata exposed by EGI DataHub.
  Moreover, OneData eases storing results without modifying simulation/processing codes, as well as maintaining usable replicas around the world.
  In this sense, EOSC is only providing the coordinating head of the service (the OneZone) and the code.
  Research communities (such as LAGO) have the freedom to self-organise and deploy the storage nodes (OneProviders) following their own needs.
  \item \textbf{EGI FedCloud} and tools for building virtual infrastructures (\textbf{IM} and \textbf{EC3}), a federation of heterogeneous IaaS providers that researchers can benefit from.
  However, every provider offers its services ``as they are'', being difficult to be directly used.
  To tackle this issue, researchers can use the EOSC services for building pre-configured virtual infrastructures.
  The selected ones for LAGO are the IM, for creating independent clusters in each IaaS provider, and EC3 for elastically deploying clusters among several providers.
  \item \textbf{B2HANDLE} is a web service able to manage a huge amount of PiDs in Handle.net for large research experiments.
  Compatible with OneData (i.e., DataHub), they allow automatically requesting PiDs for every checked catalogue to be properly referenced.
  Additionally, DataHub enables a specific web page for external users access to each catalogue and exposes the metadata to the World, to be harvested by external actors.
  One time the data is exposed, it is considered as published.
  \item \textbf{B2FIND} it is a simplified fork of a CKAN repository.
  It is designed to harvest and organise scientific metadata that was already published.
  Furthermore, it allows external users diving into the catalogues of virtual organisations thanks its search engine.
  Thus, B2FIND will act as one of the public virtual observatory in LAGO, because we do not discard to additionally use other CKAN repositories in the future, such as the ones used in the EU Joint Research Centres and other government repositories, to completely benefit from the linked-metadata in JSON-LD + DCAT2-AP format.
\end{itemize}

\subsection{Architecture for Simulations}\label{subsec:architecture-for-simulations}

The architecture follows a basic design: core intelligence of simulations are packed in Docker images, being able to automatically check, store and publish the results in DataHub, with enough metadata to be referenced by PiDs and used by harvesters, such as B2FIND, acting as virtual observatories.
As the computation is self-contained in the Docker image, the production can be easily performed on virtual infrastructures deployed by services such as EC3/IM on cloud resources, or even manually in private clusters.

\begin{figure}[!ht]
  \centering
  \includegraphics[width=0.96\textwidth]{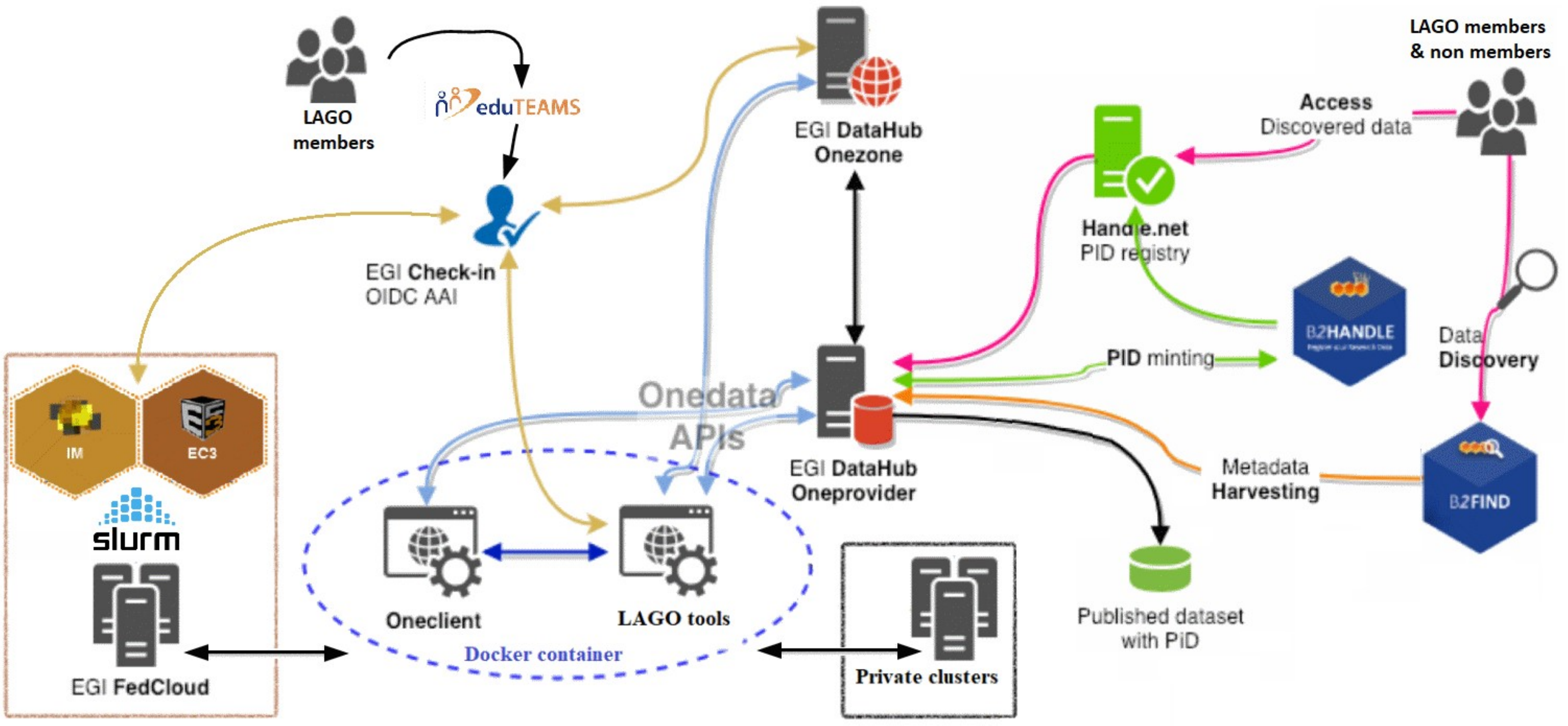}
  \caption{The architecture of the LAGO computation showing the large variety of used resources and the complexity of the cloud based implementation of the LAGO simulation services.}
  \label{fig:figlagoarch}
\end{figure}

From a user’s point of view, the architecture is shown in Figure~\ref{fig:figlagoarch}.
A researcher can access a whole data set and other additional computational resources by logging in the LAGO VO, managed by eduTEAMs. The user can create or transform old data, but uploading data to DataHub is restricted only when done via the official Docker images and software.
Thus, simulations are arbitrarily performed by researchers running the Dockers in EGI Fedcloud providers.
To perform these tasks, they dynamically deploy individual virtual machines or batch clusters through IM or EC3 services.
Additionally, they can run these Dockers on any cluster outside EOSC or inspect their content, easing the bug resolution.
Note that the only condition is to use the official Docker images and to register the results in DataHub.
This is done to ensure the correct generation of the standardised metadata for every catalogue and datasets, which will enable the publication, searches and data-mining in a later stage.
One of the main contributions of this design is how the metadata are handled and how LAGO tools are migrated into Dockers.
Just focusing on simulations, the Docker instance, named\,\cite{OnedataSim} 
encapsulates the ARTI and CORSIKA software, generating the data and metadata and storing them in DataHub.
Although researchers can create any kind of cluster, only the utilisation of Slurm workload manager is documented for now in LAGO, because it is commonly used by collaborators.
OnedataSim is already available for the LAGO community.

The integration of DataHub with B2HANDLE and B2FIND is now under testing in LAGO. These services only works together with old and reduced metadata schemes (i.e., RDF/Dublin Core), and although their compatibility with DCAT2 is scheduled by their developers, we cannot wait for them.
However, OnedataSim can additionally generate a simplified version of metadata that provisionally achieve the requirements of these services, which will be provided in the incoming months.
This will quickly enable a simplified virtual observatory for LAGO, but thanks to the mentioned design, it will not limit the interaction with other more advanced in the future.

\section{Pre-Challenge} 
\label{sec:capabilities-and-results}

For testing the current implementation of this framework, a pre-production challenge was completed.
This test involves the usage of a constrained set of cloud resources provisioned by CETA-CIEMAT, and using only the ARTI GDAS+MAGCOS+CORSIKA stages, i.e., generating S0 + S1 levels of data and metadata.

For the first time, it has been calculated the estimation of the complete $1$-day flux with no low secondary energy cutoff imposed at the stable and deploying LAGO sites described in Figure\ref{fig:figlagomap}.
Even more, we could extend the integration time to several days at some selected sites.
Note that until now, 1 to 6 hours of integration time was established as reference in LAGO simulations due to the lack of computational resources in Latin America.
Researchers had to perform several simulation runs to increase the amount of statistical data, and then, to manually combine and correct the S0-S1 outputs.
Since now, given the stochastic nature of these calculations, as the integration of the flux is increased, the impact of the statistical fluctuations is reduced, and correspondingly, it is enlarged their statistical significance and their physical relevance, besides the added easiness of their generation for publication.

The pre-challenge spent 2.7 days in a virtual cluster based on a Slurm batch system, counting on a master and 10 nodes, with 16 Intel Xeon E7 Haswell @ 2.3GHz cores, 128 GB of memory and 200 GB of disk.
Note that Docker can share both spaces unified or equally, being the needs of disk more important for the S0 generation.
In this sense, OnedataSim concurrently runs one CORSIKA task per processor, requiring less than 8 GB of RAM in total, but every task produces 5--20 GB of temporally files, thus the simulation needs 320 GB of scratched space.
However, compressed results for S0 stage are taking up over 1 TB at the OneProviders (also managed by CETA-CIEMAT) in EGI DataHub.
For the S1 stage, OnedataSim also creates a task per core, which take again the S0 data directly from DataHub and pipelines the decompression and processing, adding only a few GBs of storage data to DataHub.
The only requirement is the availability of enough memory to buffer partial downloads, which is adequately supplied by the provided nodes.

Regarding the underlying physics, the number of primaries was calculated as explained in section~\ref{sec:sc_lago} and detailed in~\cite{Asorey2018}, by setting the time integration parameter $t$ to 24 hours at the 23 LAGO selected (the stable and deploying) sites, for an idealised detector of $1$\,m$^2$, for protons to irons ($1\leq Z \leq 26$), where $Z$ is the primary atomic number.
The flux was calculated for a full hemisphere sky above the detector, 
in the energy range $\min(R_C(\phi,\lambda,\theta,\varphi)) \times Z \leq E_p / \mathrm{GeV} \leq 10^6$, where $(\theta,\varphi)$ are the primary arrival direction, $E_p$ is the primary energy, and $R_C$ is the local rigidity cutoff tensor, depending on $(\theta,\varphi)$ and also on the geographic coordinates of the site $(\phi,\lambda).$
Given the hard spectrum mentioned in equation (\ref{eq:equation}), with typical values for $\alpha(E,Z)$ varying between $-2.7$ and $-3.1$ in this energy range, the total number of primaries strongly depends on $\min(R_C)$.

For this test we only considered the secular, long term variation of the geomagnetic field, but short term variations, such as the disturbances produced by, e.g., solar activity, can be easily accounted for with the available data.

\begin{figure}[!ht]
  \centering
  \includegraphics[width=0.45\textwidth]{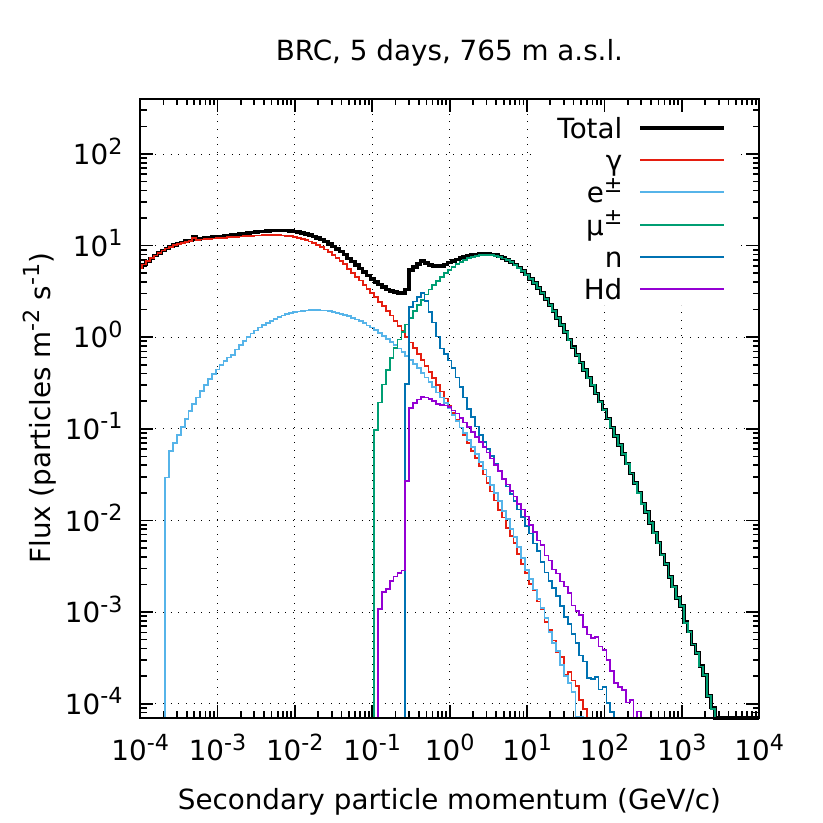}\includegraphics[width=0.45\textwidth]{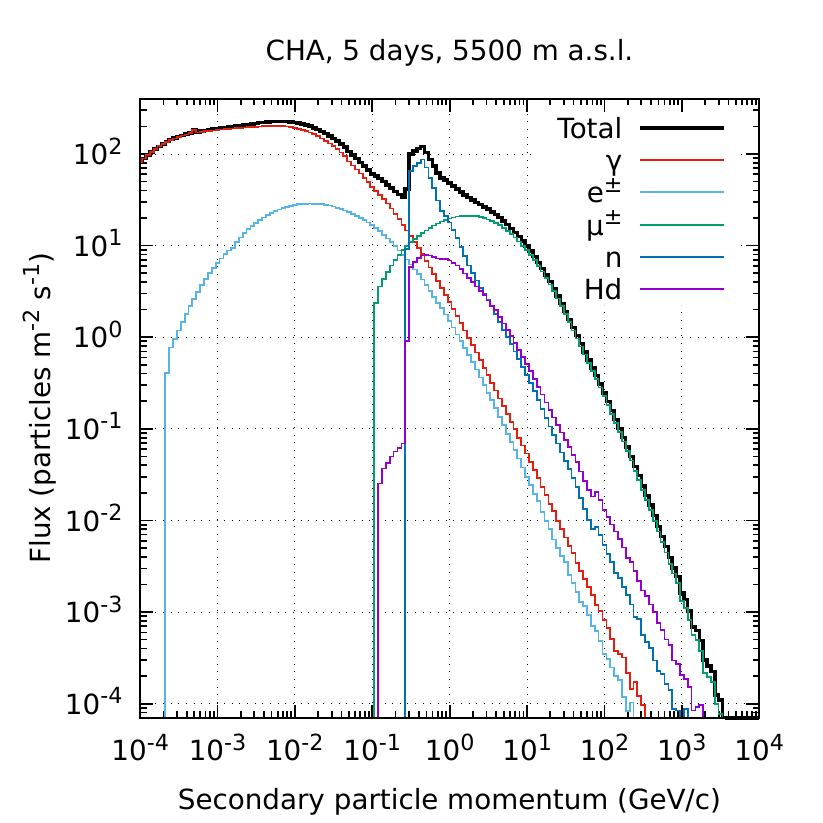}
  \caption{The energy spectrum of the flux of secondary particles expected at two of the 23 studied LAGO sites, BRC (San Carlos de Bariloche, Argentina, $765$\,m a.s.l,) is shown at left; and CHA (Chacaltaya, Bolivia, $5500$\,m a.s.l.), at right.
  It is clearly visible the impact of the atmospheric reaction at these very different altitudes. The flux is dominated by different types of particles: while the electromagnetic component dominates at low energies, the muons are present at larger momentum, even reaching the TeV scale. For these two sites, extended flux calculations were performed and the statistical significance of these results was largely increased thanks to this new ARTI cloud-based implementation.}
  \label{fig:pre-challenge}
\end{figure}

In the Figure~\ref{fig:pre-challenge} the obtained secondary spectra are shown for two sites: San Carlos de Bariloche (BRC), Argentina, $765$\,m a.s.l. (left), and Chacaltaya (CHA), Bolivia, $5500$\,m a.s.l.\
The impact of the atmospheric reaction is visible for these sites, accounting for the absorption of secondary particles in the air due to their large altitude difference.
Given the importance of these two sites for our observatory, an extended integration times of $5$\,days of flux was simulated to increase the statistical significance of the results.

These spectra are used to evaluate the expected flux of secondaries and determine, e.g., the optimal detector geometry, the detector calibration parameters, or the typical energies expected for the different components of the atmospheric radiation background.
Even more, the presence and the behaviour of the neutron peak in the flux motivate the recent evaluation of increasing the neutron sensitivity for space weather studies by doping the water with NaCl\,\cite{Sidelnik2020}.

\section{Conclusions and Future Work}\label{sec:conclusions}

The continuous evolution of high energy and astroparticle physics demands an increasing amount of computational resources to achieve more accurate results.
Vast amounts of data and metadata are produced every day around the world, which should be effectively accessed, preserved and curated.
While the computing facilities continue increasing their processing and storing capabilities, the need to standardise the calculation methods and access to resources is still in progress at several scientific communities.

Decentralised observatories and experiments, such as LAGO, should ensure the Open Science paradigm to take full advantage of their collaborative approach.
Therefore, besides the integration of services from EOSC and the creation of a new virtual organisation, the core contributions to the standardisation of LAGO computation are focused on the definition of data and their metadata, as well as their generation and storage by tools deployed in Docker images.
In this work, we show a first glimpse of how ARTI, the LAGO simulation sequence, has been successfully adapted to this framework.
Promising results have been obtained and the accuracy and statistical significance can be easily improved by orders of magnitude if more resources from EOSC were used.

Therefore, other computational challenges will be scheduled during 2021, involving the full calculation for the whole LAGO sites, other locations of physical interest, including different site altitudes, muography experiment locations and underground laboratories around the World~\cite{Rubio-Montero2021icrc}.

Even more, for some planned uses of this framework, such as the study of high energetic secondary muons at the tens of TeV scale, the integration times needed could be of at least 1-year of simulated flux, requiring a computational power several times higher than the needed for the calculations presented in this work.

Regarding Open Science, the next step will be coupling the framework to metadata harvesters to enable them as the gateways of the virtual observatory of LAGO data.

\section*{Acknowledgements}

The LAGO Collaboration is thankful to the Pierre Auger Collaboration for their continuous support.
It is also acknowledged the support of the LAGO Collaboration members.
This work was carried out within the 'European Open Science Cloud-Expanding Capacities by building Capabilities' (EOSC-SYNERGY) project, founded by the European Commission’ Horizon 2020 RI Programme under Grant Agreement nº 857647.
HA thanks to Rafael Mayo García and CIEMAT for their warm welcome and support at Spain.

\bibliographystyle{unsrt}  
\bibliography{lago-wsc21}
\end{document}